\documentclass[10pt]{iopart}
\usepackage{psfig}
\newcommand{\be}{\begin{equation}}
\newcommand{\ee}{\end{equation}}
\newcommand{\D}{\Delta}
\newcommand{\bea}{\begin{eqnarray}}
\newcommand{\eea}{\end{eqnarray}}

\begin{document}

\title[Quarkonium at collider energies]{Suppression of Quarkonium Production
in Heavy Ion Collisions at RHIC and LHC}

\author{L. Gerland$\sharp$, L. Frankfurt$\sharp$\ddag, M. Strikman\S,
H. St\"ocker$\sharp$, W. Greiner$\sharp$}
\address{$\sharp$ Institut f\"ur Theoretische Physik der 
J.W.Goethe-Universit\"at\\
Robert-Mayer-Str. 8-10, D-60054 Frankfurt a.M., Germany}
\address{\ddag Tel Aviv University, Tel Aviv, Israel}
\address{\S Department of Physics, Pennsylvania State University\\
University Park, PA 16802, USA}

\begin{abstract}
A model for the production of quarkonium states
in the midrapidity region at RHIC and LHC energy range is presented
which explores well understood properties of QCD only.
An increase of the quarkonium hadronisation time with the initial energy 
leads to a gradual change 
of the most important phenomena from fixed target- to collider-energies.
We evaluate nuclear effects in the quarkonium production 
due to medium modification of the momentum distribution of the 
heavy quarks produced in the hard interactions,
i.e. due to the broadening of the transverse 
momentum distribution. Other nuclear effects, i.e. nuclear 
shadowing and parton energy loss, are also evaluated.
\end{abstract}

\section{Introduction}
At fixed target energies the hadronisation time $t_f$
for $Q\bar Q$ pair
is small as compared to the nuclear radii, at collider energies
this time is much larger than the size of the nucleus.
Thus, at fixed target energies, the interaction of heavy quarkonium with the 
surrounding medium can be described effectively as 
hadronic~\cite{farrar, gerland} 
even within  the hadronisation time.
However, at collider energies a description of the process in terms of hadron 
degrees of freedom becomes questionable since a bound state
of the heavy quark pair is formed far outside of the nucleus. 

One major effect at collider energies is the transverse 
momentum broadening of $Q\bar Q$ pairs travelling through a medium.
This changes the overlap integral
between the $Q\bar Q$ pair and the wave functions of the heavy quarkonia.
It is important that soft interactions of heavy quarks with nucleons 
do not change the longitudinal momentum of heavy quarks  which may appear 
sensitive to the interaction with comovers which are neglected in this paper.
Nuclear effects in the momentum distribution within the heavy quark pair 
due to the radiation of gluons which is sensitive to 
the properties of the QCD-matter produced in a nucleus-nucleus 
collision at collider energies~\cite{baier}
will be estimated separately in the section six.

The typical time scales for the creation of a hidden heavy flavour
state are investigated in section two, in the third section 
a model is developed to calculate the influence of nuclear effects
on the production of states with hidden heavy flavor in the midrapidity region
at collider energies. In the fourth section this model 
will be used to predict heavy quarkonium products in $AB$ collisions 
at RHIC and LHC and in section
five and six quarkonium suppression due to nuclear effects on the 
parton distribution functions and
parton energy loss are evaluated.

\section{Coherence Lengths}
There are different time scales relevant for the 
production of quarkonium states: \\
1) the time needed 
to produce a heavy quark pair in a hard collision, \\
2) the time needed for a $Q\bar Q$ pair to form a bound state.

The production time of a $Q\bar Q$ pair in its rest frame is given by
$\tau_p={1\over m_Q}$. This is 0.13~fm/c for $c\bar c$ and 
0.05~fm/c for $b\bar b$. The Lorentz factor of the pair at 
midrapidity in the rest 
frame of the target is $\gamma={\sqrt{s}\over 2m_N}\approx 10$, 
20, 100 and 3000 for a 
Quarkonium state in the midrapidity region at
SPS, Fermilab, RHIC and LHC energies. 
Thus, at SPS and Fermilab fixed target energies $\gamma c\tau_p$ is 
smaller than
the average internucleon distance in nuclei 
$r_{NN}\approx 1.8$ fm.
Thus, the production of heavy quark pairs is incoherent.
However, at RHIC the production distance of $c\bar c$ 
pairs is already as large as the diameter of a gold nucleus,
and for $b\bar b$ pairs $c\tau_p>1.8$ fm, but this is still small as 
compared to the nuclear radius. At LHC both production distances 
exceed the diameter of a lead nucleus by an order
of magnitude. 

The hadronisation time $t_{H}$ resp. the coherence length $l_c$
of heavy Quarkonium is
$l_c=c\cdot t_{H}={1\over \Delta E}\approx{\gamma\over \D M}$
with:\\
$\Delta E=\sqrt{p^2+(M_{Q\bar Q}+\Delta M)^2}-\sqrt{p^2+M^2_{Q\bar Q}}
\approx {(M_{Q\bar Q}+\Delta M)^2-M^2_{Q\bar Q}\over 2p}\approx
{M_{Q\bar Q}\Delta M\over p}={\Delta M\over \gamma}\; .$\\
Here $p$ is the momentum of the Quarkonium in the rest frame of the target,
$M_{Q\bar Q}$ is the mass of the $Q\bar Q$-pair and
$\Delta M={\int \psi^2(k){k^2\over M_{Q}}{\rm d}^3k/
\int \psi^2(k){\rm d}^3k}\;$,
where $\psi(k)$ is the wavefunction of the Quarkonium state in momentum space.
That means $\Delta M$
is the average kinetic energy of the $Q\bar Q$-pair in the bound state
and $l_c/\gamma=0.44 (0.34)$~fm for the $J/\Psi$ ($\Upsilon$).

Thus, for charm and bottom production at RHIC and LHC
$l_c> 2\cdot R_A$ ($R_A$ is the nuclear radius),
$l_c< 2\cdot R_A$ for fixed target at Fermilab
($\gamma\approx 20$) and at SPS energies. The applicability of the
approach developed in this paper requires that $l_c> 2\cdot R_A$ which is 
fulfilled at RHIC and LHC
(and at large $x_f$ at Fermilab fixed target energy range, where 
$\gamma\sim 100$ also can be reached).

\section{Description of the Model Approach}

 We assume here that $Q\bar Q$ pairs are produced in $AB$ collisions
predominantly in hard collisions. 
The basic quantity is the cross section of
production of $Q\bar Q$ pair with light cone momenta $z_{i},k_{i}$,
which we parametrize as 
${d\sigma(AB\rightarrow Q\bar Q+X)\over {\rm d}^2 k_1 {\rm d}z_1 {\rm d}z_2 {\rm d}^2 k_2}
=D_{AB}(z_1,z_2)\cdot \exp(-B(AB)(k_1^2+k_2^2))$.
Here $k_i\;\{ i=1,2\}$ are the transverse momenta of the $Q$ and the $\bar Q$
quark and $z_i\;\{ i=1,2\}$ are the fractions of their longitudinal momenta.
Such a factorization does not contradict the data in 
pp collisions~\cite{e789.1}.

To evaluate the suppression of hidden heavy flavour
production resulting from the broadening of the transverse momentum
distributions of $Q$ quarks due to final state interaction,
we deduce first a relationship between the slopes
for the various processes of heavy quark production.
In the following we use the relative transverse momentum
$k_t={k_1-k_2\over 2}$ and the total transverse momentum
$p_t=k_1+k_2$ of the pair, writing
${d\sigma(AB\rightarrow Q\bar Q+X)\over {\rm d}^2 k_t {\rm d}^2 p_t 
{\rm d} z_1 {\rm d} z_2} 
=D_{AB}(z_1,z_2)\exp\left(-B(AB)\left(-{p_t^2\over 2}-2k_t^2\right)\right)$.

To take into account possible nuclear effects on 
the longitudinal momentum distribution we make the ansatz
$D_{AB}(z_1,z_2)=D(AB)\cdot f_{AB}(p_z)\cdot \exp\left(-{k_z^2\over C^2_{AB}}\right)\; .$
where $p_z$ and $k_z$ are the total and relative longitudinal 
momentum. We further assume that
$f_{AB}(p_z)=f_{pp}(p_z)\;$, which means that we neglect 
parton energy losses of the pair, 
this effect will be discussed later on. 
The normalization
condition follows from the QCD factorization 
theorem~\cite{bodwin} for the total cross section:
$
{D(AB)\over B(AB)^2\cdot C_{AB}}={AB\cdot D(pp)\over B(pp)^2\cdot C_{pp}}
$
The differential cross sections are proportional to the square of the two body
wave functions $\phi$. 

The production cross section of bound states of heavy quarks 
is proportional to the overlap integral of the two-body
wave function and the wave function of the bound $\psi(k_t)$ state to get
$
{{\rm d}^3\sigma(AB\rightarrow Quarkonium+X)\over {\rm d}^2 p_t {\rm d} p_z}
\propto\left|\langle\psi(k_t,k_z)|\phi_{AB}(k_t,p_t,k_z,p_z)\rangle\right|^2\; .
$
Here we neglected the difference between the 
current quark mass in the two body wave function
and the constituent quark mass in the wave function of the bound state.
With this one can evaluate the survival probability:
$
S\equiv{\sigma (A+B\rightarrow Quarkonium+X)\over
AB\cdot\sigma (p+p\rightarrow Quarkonium+X)}
$. Our final result is then
\be
S= {B(AB)C_{AB}\over B(pp)C_{pp}}
\left|\frac{\int{\rm d}^3k \psi(k) \exp(-B(AB)
k_t^2)\exp\left(-{k_z^2\over 2C^2_{AB}}\right)}
{\int{\rm d}^3k \psi(k) \exp(-B(pp) k_t^2)\exp\left(-{k_z^2\over
2 C^2_{pp}}\right)}\right|^2
\label{overlap2}
\ee
up to nuclear effects in the structure function. 

Note that if one defines the survival probability 
as the ratio of the
differential cross section ${{\rm d}^2\sigma \over {\rm d}^2 p_t}$ 
for nuclear and nucleon targets
their $p_t$ dependence would be a factor 
$\exp\left(-{B(AB)-B(pp)\over 2}p_{t}^2\right)$.
That means that the $p_t$ dependence of $J/\Psi$ suppression is 
due to the broadening of the transverse momentum distribution
as a result of the final sstate interactions of the $Q$ quarks in the 
nuclear medium. 

\section{Predictions for Collider Energies}

To adjust the parameters of eq.~(\ref{overlap2}) we use the 
quarkonium wave functions of ref.~\cite{eich,buch} and the following data: 

The E789-group (Fermilab) measured a value of
$B(pAu)=0.91\pm 0.12\; {\rm GeV}^{-2}$~\cite{e789.1} with the fit function
${{\rm d}N\over {\rm d}k_t^2}(D,\bar D)\propto \exp(-B(pAu)\cdot k_t^2)$
in proton-gold collisions for the transverse momentum distribution of $D$ 
and $\bar D$ mesons. The same group found
$n=0.55\pm 0.01\; {\rm GeV}^{-2}$~\cite{e789.2} with
${{\rm d}N\over {\rm d}p_t^2}(J/\Psi)\propto \exp(-n\cdot p_t^2)$,
also in proton-gold for the transverse momentum distribution
of $J/\Psi$'s. Within the error bars these values of $n$ and
$B(pA)$ satisfy the relation $n={B(pA)\over 2}$ 
that was predicted in the 
previous section.
With the fit~\cite{peng2} to the $p_t$ broadening of $J/\Psi$'s this yields
$B(pp)=1.18 (1.5){\rm GeV}^{-2}$. 

There are no data for the nuclear broadening of the longitudinal 
momentum distribution.
We assumed isotropic production in the $c\bar c$ rest frame which 
yields $C_{pp}^2=1/(2\cdot B(pp))\approx 0.42$ GeV${}^2$. 
In fig. 1 the result of eq.~(\ref{overlap2}) is plotted versus the
transverse momentum broadening of the $J/\Psi$'s: $\D p_t^2=2/B(pp)-2/B(AB)$.
$\D p_t^2=0.48$ GeV${}^2$ was found at Fermilab energies in $pAu$. 
$C_{AB}^2=C_{pp}^2$ is used as a first approximation. 
$S\approx 1$ for the $J/\Psi$. That means
there is practically no change due to 
the broadening of the transverse momentum
distribution. 

For the $\Upsilon$ meson we use $<p_t^2>=2.59$ GeV${}^2$~\cite{e772} and thus
$B(pp)=0.77 {\rm GeV}^{-2}$ and 
$C_{pp}^2=1/(2\cdot B(pp))\approx 0.65$ GeV${}^2$.
For the $\Upsilon$ meson production we obtain even a slight enhancement.

\parbox{7.3cm}{
\psfig{figure=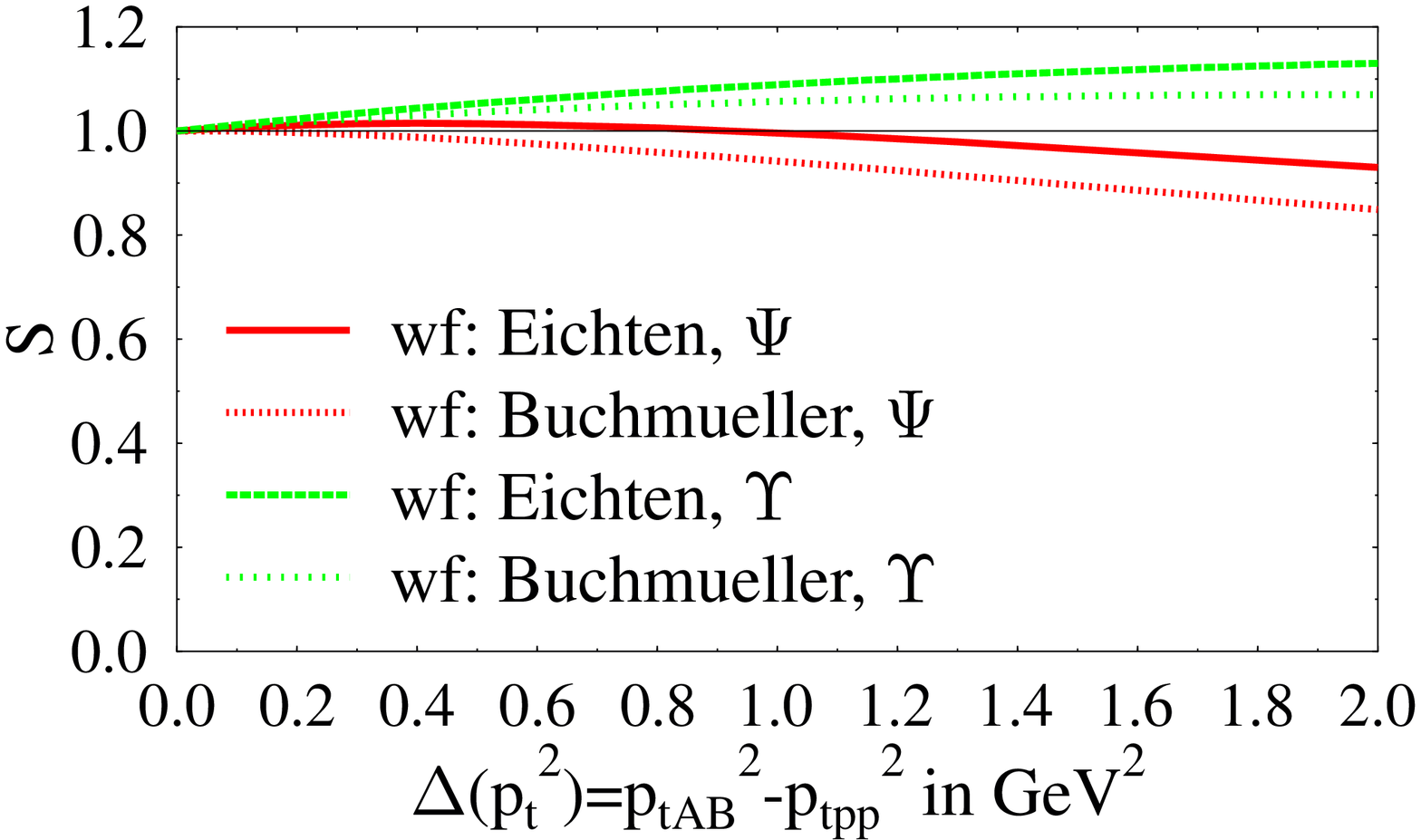,width=7.cm}
}
\parbox{5cm}{{\small Figure 1: The righthandside of eq.(\ref{overlap2})
vs. the transverse momentum broadening for the $J/\Psi$ and 
the $\Upsilon$ is plotted.}
}

Ref.~\cite{peng2}  reported that $\D p_t^2$ is nearly a factor 5 larger for
$J/\Psi$'s than for Drell Yan pair production. In our model this is
due to a color factor of ${9\over 4}$ because $c\bar c$ production
is dominated by gluon-gluon fusion, while Drell Yan is 
due to quark-antiquark 
annihilation and due to the factor of two between 
the transverse momentum distribution
of $c$ resp. $\bar c$ Quarks and the bound states evaluated above.

\section{Nuclear Effects on the Gluon Distribution Functions}

The term nuclear effects on the gluon distribution functions means that the
ratio of the gluon distribution functions 
in a nucleus $A$ and a nucleon $N$ differs from 1, i.e
${G_A(x,Q^2)\over AG_N(x,Q^2)}<1$ in the shadowing region,
${G_A(x,Q^2)\over AG_N(x,Q^2)}>1$ 
for $0.1\le x\le 0.3$ and for $x>0.8$. 
The Bjorken $x$'s of interest for the production of $J/\Psi$'s $x=0.0155$
and $\Upsilon$'s $x=0.0473$ are $x(Q\bar Q)={M_{Q\bar Q}\over \sqrt{s}}$.

Up to now, the transition points between these regions are not well
known. Especially for gluons, the error bars of the data are large.
Also from a theoretical point of view they are not unambigious. Therefore,
the calculation of these nuclear effects, at not too small $x$, is model 
dependent. To reduce this 
model dependence we use here two different recent approaches from
ref.~\cite{eskola1,eskola2} (called $EKS$ in the following) and from
ref.~\cite{FS} ($FS$). (Nuclear shadowing in ($FS$) has been 
calculated in a nearly
model independent way, but the region of enhancement 
requires modelling.)

In tab. 1 one can see that quarkonium production at RHIC
is enhanced while it is shadowed at LHC.
At LHC energies the predictions
of both approaches become qualitative different. 
The latter one based on significantly larger 
cross section of diffraction in the gluon channel 
or equivalently on the $9/4$ enhancement of 
color dipole build of color octets.
It predicts a stronger shadowing of gluons
than the shadowing of quarks and of antiquarks, while in the $EKS$ model 
it is as strong as the shadowing for sea-quarks. 

\section{Suppression due to Parton Energy Loss}

Parton energy loss shifts the partons to smaller x.
We use here the parton energy loss model reviewed in ref.~\cite{baier}.
$\Delta E$ is the parton energy loss in a medium.
For cold QCD-matter $\Delta E=-4 \mbox{ GeV}\cdot 
\left({L\over 10 \mbox{ fm}}\right)^2$.
For hot QCD-matter $\Delta E=-60 \mbox{ GeV}\cdot 
\left({L\over 10 \mbox{ fm}}\right)^2$.

With $L= R_{Au}=7$ fm this yields
$\Delta E=-2 \mbox{ GeV}$ for cold matter and
$\Delta E=-30 \mbox{ GeV}$ for hot matter.
For RHIC energies this value has to be divided by 100
since in~\cite{baier} the parton energy loss was calculated in the 
rest frame of the target
while we are interested in the midrapidity
region.
Therefore, an initial parton with
$x(J/\Psi)=0.017$ in hot QCD-matter
is needed to produce a $J/\Psi$.   
For $pA$ collisions\\
\parbox{7.3cm}{
\begin{tabular}{|c|c|c|c|c|}
\hline
$\sqrt{s}$ in GeV & 60 & 140 & 200 & 6000 \\
\hline
$J/\Psi$ (FS) & 1.11 & 1.05 & 1 & 0.65\\
\hline
$J/\Psi$ (EKS) & 1.16 & 1.01 & 0.96 & 0.84\\
\hline
$\Upsilon$ (FS) & 1.09 & 1.04 & 1.09 & 0.83\\
\hline
$\Upsilon$ (EKS)  & 1.104 & 1.138 & 1.12 & 0.87\\
\hline
\end{tabular}}
\parbox{5cm}{{\small Table 1:
${G_A(x,Q^2)\over AG_N(x,Q^2)}$ at different initial energies (see text).}}\\
this yields an suppression factor of
${G(0.017,3.1\mbox{ GeV})\over G(0.0155,3.1\mbox{ GeV})}=0.87$
for the gluon distribution parametrization of Ref.~\cite{cteq4,cteq42}. For the $\Upsilon$ 
this factor is 0.94.
For cold QCD-matter this is negligible, since $x=0.0156$ for the $J/\Psi$.
This yields a suppression factor of 0.99.
For $AB$ collisions this values have to be squared, since both initial gluons
can loose energy. Therefore, the suppression for the $J/\Psi$ ($\Upsilon$)
in hot matter in $AB$ collisions is 0.76 (0.88).
This educated guess must suffice for the time being, because the parton 
energy loss is theoretical not
well understood. 

\section{Discussion and Conclusion}

\begin{itemize}
\item
the relevant physics changes due to coherence length:\\
$\longrightarrow$
hadronic description at fixed target,
partonic at colliders
\item
nuclear effects on the parton distribution functions yields:\\
$\longrightarrow$
anti-shadowing at RHIC,
shadowing at LHC
\item
parton energy loss seems to be uneffective in cold QCD matter
\item
other possible effects not discussed here include:\\
$\longrightarrow$
interaction with comovers\\
$\longrightarrow$
long living phase of hot QCD-matter\\
$\longrightarrow$
$J/\Psi$-production and $\Upsilon$ comes also from the decay of higher
resonances
\item
$pA$ collisions are needed to distinguish between the different effects.
\end{itemize}

\ack
The authors wants to thank V. Guzey for discussions and technical support 
concerning nuclear effects in the parton distribution functions. 
LG wants to thank the Josef Buchmann Stiftung. LF acknowledges support by
the Alexander von Humboldt Foundation.

\end{document}